\documentclass[aps,pra,twocolumn,showpacs,floatfix]{revtex4}
\newcommand{\fmod}{\ensuremath{f_{\textrm{\footnotesize{mod}}}}}
\newcommand{\tmod}{\ensuremath{T_{\textrm{\footnotesize{mod}}}}}
\newcommand{\sinc}{\ensuremath{\textrm{sinc}}}
\usepackage{graphicx,color}
\usepackage{epsfig}
\usepackage[T1]{fontenc}
\usepackage[latin1]{inputenc}
\usepackage{alltt}
\usepackage{amssymb}
\usepackage{amscd}

\begin{document}
\setlength{\unitlength}{1mm}
\bibliographystyle{unsrt}
\title{An experimental study of intermodulation effects\\
 in an atomic fountain frequency standard}
 \author{J.
Gu\'ena\footnote{On leave from Laboratoire Kastler Brossel, CNRS UMR 8552, 24 Rue Lhomond, F-75231
Paris Cedex 05, France}, G. Dudle, P. Thomann\footnote{LTF-IMT, Universit\'e de Neuch\^atel, rue
A.-L. Breguet 1, CH-2000 Neuch\^atel, Switzerland}} \affiliation{ METAS, Lindenweg 50, 3003
Bern-Wabern, Switzerland}
\date{Februar 19, 2007}
 \begin{abstract}
The short-term stability of passive atomic frequency standards, especially in pulsed operation, is
often limited by local oscillator noise \textit{via} intermodulation effects. We present an
experimental demonstration of the intermodulation effect on the frequency stability of a
{\it{continuous}} atomic fountain clock where, under normal operating conditions, it is usually too
small to observe. To achieve this, we deliberately degrade the phase stability of the microwave
field interrogating the clock transition. We measure the frequency stability of the locked,
commercial-grade local oscillator, for two modulation schemes of the microwave field: square-wave
phase modulation and square-wave frequency modulation. We observe a degradation of the stability
whose dependence with the modulation frequency reproduces the theoretical predictions for the
intermodulation effect. In particular no observable degradation occurs when this frequency equals
the Ramsey linewidth. Additionally we show that, without added phase noise, the frequency
instability presently equal to 2$\times$10$^{-13}$ at 1s, is limited by atomic shot-noise and
therefore could be reduced were the atomic flux increased.
\end{abstract}

\maketitle
%\author{J. Gu\'ena\thanks{Present address: LNE-SYRTE,
%Observatoire de Paris, 61 Avenue de l'Observatoire, 75014 Paris, France}, \\
%}

\section{Introduction}
\label{sec:introduction} It is well established that the ultimate short-term stability of pulsed
passive frequency standards can be limited by intermodulation effects, also known as the Dick
effect \cite{{dick87},{dick90}}. First described in the case of ion traps in 1987, the effect has
regained interest with the development of cold cesium fountains \cite{santarelli98}, since most of
these devices are operated with a pulsed scheme. The Dick effect arises from down-conversion of
local oscillator frequency noise at even harmonics of the cycle rate into the fundamental frequency
band of the locking loop, thereby degrading the achievable short-term stability. In a pulsed
fountain one can make the effect negligible by employing a local oscillator exhibiting an
ultra-high stability, such as a cryogenic sapphire oscillator \cite{{santarelli99}, {mann2001}}. An
alternative route would be to ensure that at all times there are atoms between the two Ramsey
pulses. One way to achieve this is to use a multipulse or a juggling atomic fountain
{\cite{oshima95},\cite{fertig99}. The way we are currently pursuing is the \textbf{continuous}
fountain approach. Rather than launching atoms in successive clouds, a such fountain produces a
continuous beam of laser-cooled atoms.

Theoretical investigations, presented in a previous communication \cite{AJo1}, have shown that
intermodulation effects can also exist in a continuous fountain, albeit at a much lower level. But
contrary to pulsed standards, they can be cancelled completely in the case of a continuous fountain
if the modulation frequency of the interrogation is carefully selected; namely, if this frequency
is an \textit{odd} harmonic of the linewidth of the Ramsey resonance, the intermodulation effect
cancels for any modulation scheme. In this paper we present an experimental verification of the
suppression of the intermodulation effect in a continuous fountain when this condition is fulfilled
and observe the degradation of the short-term stability when it is not. Results are compared with
the theoretical prediction.

In Section \ref{sec:setup}, we present the experimental set-up used for this work, including a
brief description of our continuous fountain, dubbed FOCS-1, and the frequency control loop.
Section \ref{sec:perfos} is devoted to the present performances of FOCS-1 in terms of flux,
signal-to-noise ratio and stability. In Section \ref{sec:dick}, we summarize the results of the
model developed in \cite{AJo1} and use them to predict the intermodulation effect in a realistic
experiment. Measurements and results are presented in Section \ref{sec:measurements}.

\section{The continuous fountain clock set-up and interrogation method}\label{sec:setup}

\subsection{Fountain set-up}\label{sec:focs1}

 Fig. \ref {fig:setup} shows a sketch of FOCS-1.
 A continuous 3D optical molasses (OM)
 captures atoms from the low velocity tail of a Cs
 vapor in an ultra-high vacuum chamber. The cold atoms
 are launched
 upwards with a longitudinal velocity of $\simeq$~3.8~m/s using
 a moving molasses scheme (up- and down-going beams at 45$^{\circ}$ from the vertical, detuned by
 $\pm$~3.2~MHz). The cold atom beam (isotropic temperature of about 60~$\mu$K) is tranversally collimated by two orthogonal 1D optical
 molasses, which reduces the transverse temperature to about 7~$\mu$K. The cooling beams are tuned to the
 cross-over 43'-45' of the D2 line, \textit{i.e.} 25 MHz to the blue side of the 44'.
 A repumper beam tuned to 34' is superposed so that all atoms are cooled in the F=4 ground state hyperfine level.

In the continuous fountain, the two Rabi interactions must be spatially separated. For this reason
the transverse cooling plane is tilted by about 1.8$^{\circ}$ from horizontal and the atoms
describe an open parabola with a flight duration T$\simeq$~0.5~s in between the two Rabi
interaction zones (yielding a Ramsey fringe of FWHM $\simeq$~1~Hz). In the present set-up the atoms
in F=4 are partially pumped into F=3 by the fluorescence light scattered by the OM source. Since
this depumping is unavoidable, we have decided to complete the transfer into F=3 by a transverse
laser beam above the cooling plane resonant for 44'. The remaining atoms in F=4, m$_{F}\neq0$
represent typically 40\% of the population inversion of the clock transition F=3, m$_{F}$=0
$\rightarrow$ F=4, m$_{F}$=0 and contribute noise.

The transit time in each Ramsey zone is of order 10~ms. A C-field B $\thickapprox$~70~nT defines
the vertical quantisation axis and lifts the degeneracy of the F=3, m$_{F}$ $\rightarrow$ F=4,
m$_{F}$ transitions. After the second Rabi interaction, atoms in F=4 only are detected by induced
fluorescence on the F=4 $\rightarrow$ F'=5 cycling transition. The detection efficiency limited by
the solid angle of collection is a few percent.
 With a probe power 1.5~mW and waist diameter 10~mm, the number of photons detected per atom
is $\gg$ 1, so that the detected shot noise is limited by atomic shot noise. The photodetector
(Hamamatsu) with a feedback resistance 1~G$\Omega$, has low dark noise
($4\times10^{-15}$~A~Hz$^{-1/2}$). The output of the current-to-voltage converter is further
amplified by a factor of 100.

\begin{figure}[thb]
\begin{center}
\includegraphics[width=7cm]{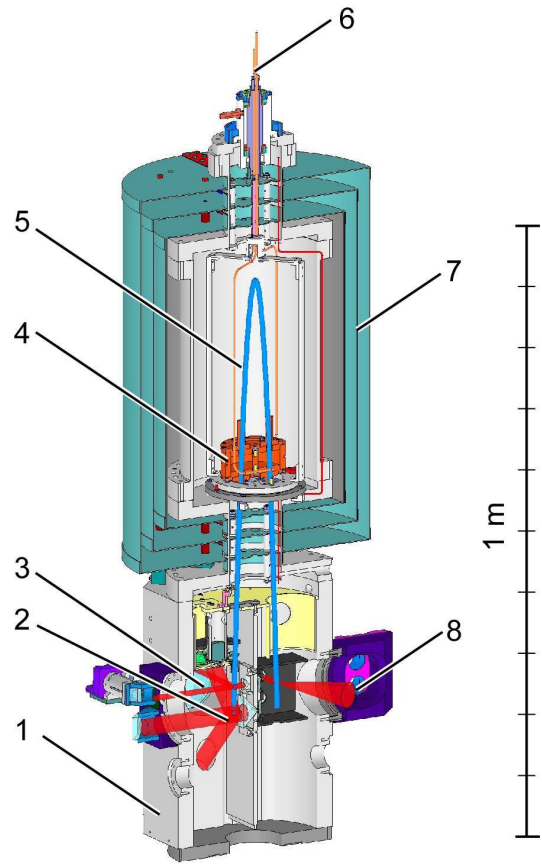}
\caption{Fountain set-up. 1: UHV chamber; 2: optical molasses capture region; 3: transverse
cooling; 4: microwave cavity; 5: parabolic flight; 6: microwave field feedthrough; 7: magnetic
shielding; 8: probe detection beam.} \label{fig:setup}
\end{center}
\end{figure}

\subsection{Continuous Ramsey interrogation}\label{sec:loop}

A block diagram of the frequency control loop is shown in Fig.~\ref{fig:loop}. We start from a
commercial voltage-controlled quartz crystal oscillator VCXO (BVA 8607 from Oscilloquartz) which
displays an Allan deviation of $10^{-13}$ up to 100~s. A frequency synthesiser upconverts the
10~MHz output of the VCXO to 9180~MHz with a phase modulation at 12.6~MHz provided by an external
generator (HP3325 synthesiser). This generates the microwave field with a carrier frequency at the
clock transition frequency (9 192 631 770~Hz) and an amplitude adjusted to induce $\pi/2$ Rabi
pulses in the Ramsey cavity. To produce an error signal for the frequency corrections, the
\textit{phase} of the 12.6~MHz oscillation issued from the HP3325 synthesiser is square-wave
modulated with a p.p. amplitude of $\pi/2$. The phase of the microwave field at the clock frequency
is thus modulated with the same waveform and amplitude. The waveform, frequency \fmod~ and
amplitude of the phase modulation are controlled by the reference output of a digital lock-in
amplifier (DLA). The fluorescence photodetector signal is square-wave demodulated in the DLA,
integrated and the correction voltage applied to the VCXO. The locking loop is controlled by a
Labview software routine.

\begin{figure*}[thb]
\begin{center}
\resizebox{0.75\textwidth}{!}{%
\includegraphics{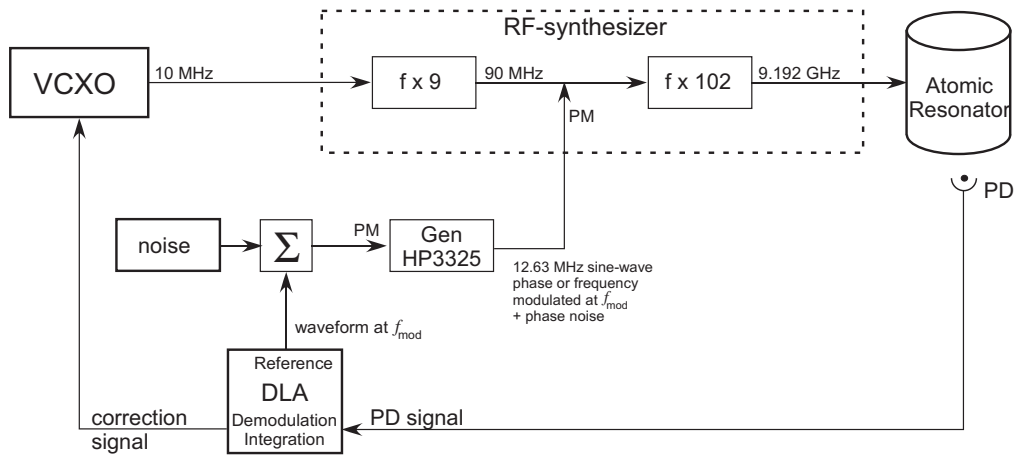}
} \caption{Block diagram of the frequency control loop. VCXO: voltage-controlled quartz crystal
oscillator; PD: fluorescence photodetector. DLA: digital lockin amplifier. Gen: RF synthesiser. PM:
phase modulation. The use of the noise generator (noise) and summing amplifier $\Sigma$ is detailed
in Sect. \ref{sec:measurements}. }\label{fig:loop}
\end{center}
\end{figure*}

\section{Signal-to-noise and present clock stability}\label{sec:perfos}

%\subsection{Signal-to-noise}\label{sec:snr}
In a continuous fountain, if all technical noise sources can be suppressed sufficiently, the
achievable stability is limited only by the signal-to-noise ratio $S/N$ of the detected atomic flux
which itself scales with the square root of this flux (atomic shot noise limit). The Allan
deviation $\sigma_{y}$, a measure of the frequency instability of the fountain, is then expected to
scale with the inverse square root of the atomic flux, $\sigma_{y} \propto \phi^{-1/2}$. To verify
that our fountain satisfies these conditions, we have first measured the noise as a function of the
atomic flux.

For these measurements, the atomic flux was varied in two independent ways, either i) by changing
the power of the repumper laser or ii) by modifying the power of the cooling beams of the
3D-optical molasses. The fluorescence photodetector signal was spectrum analysed (Stanford Research
Systems SR760 FFT). The noise was observed to be white between 0.1~Hz and 20~Hz with a dominant
contribution from atomic fluorescence, as can be seen in an example of a recording shown in Fig.
\ref{fig:snr}a. Figure \ref{fig:snr}b displays the measured noise as a function of the detected
atomic flux for both methods. The small noise contribution from the background signal due to
scattered probe light
 ($\sim~5\times10^{-15}$~A~Hz$^{-1/2}$) has been  subtracted quadratically.
The experimental data are fitted with a power law $N~=~S^k$ where $N$ represents the noise and $S$
the atomic signal. The value obtained for $k$ lies $\sim 2.5\sigma$ above the value 0.5 expected
for pure atomic shot noise.

From the signal-to-noise ratio one can also deduce the number of detected atoms. Indeed, it can
readily be shown that the atomic flux is related to the signal-to-noise ratio by $\phi = 2 \left( S
/ N \right)^2$.

Using this equation, we infer that for the data point corresponding to the strongest atomic signal,
the useful flux \footnote{This is the flux of atoms having undergone the clock transition excluding
the background atoms in F=4, m$_{F}\neq0$.} is of order $2\times10^5$ at/s.

\begin{figure}[thb]
\begin{center}
\includegraphics[width=8cm]{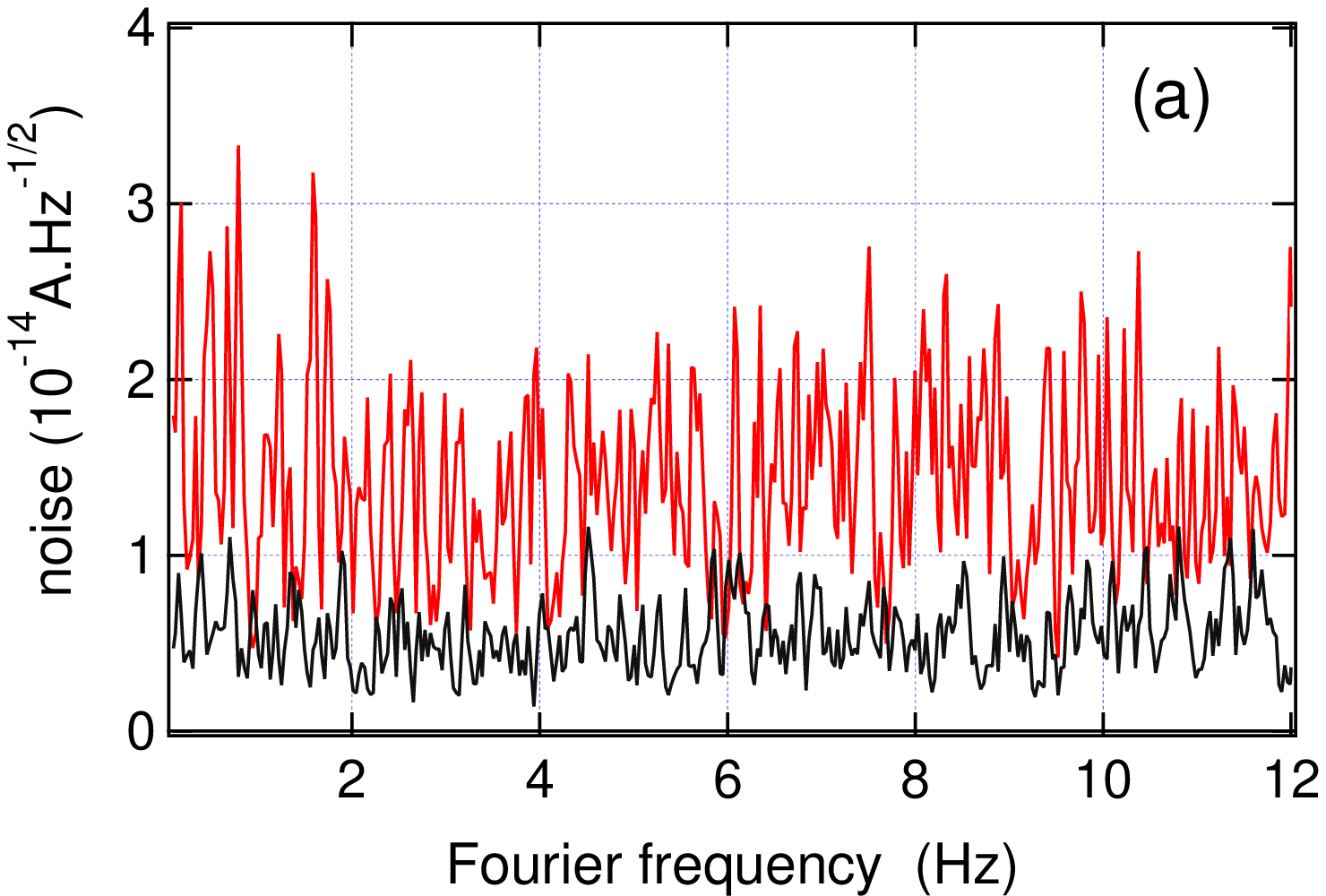}
\includegraphics[width=8cm]{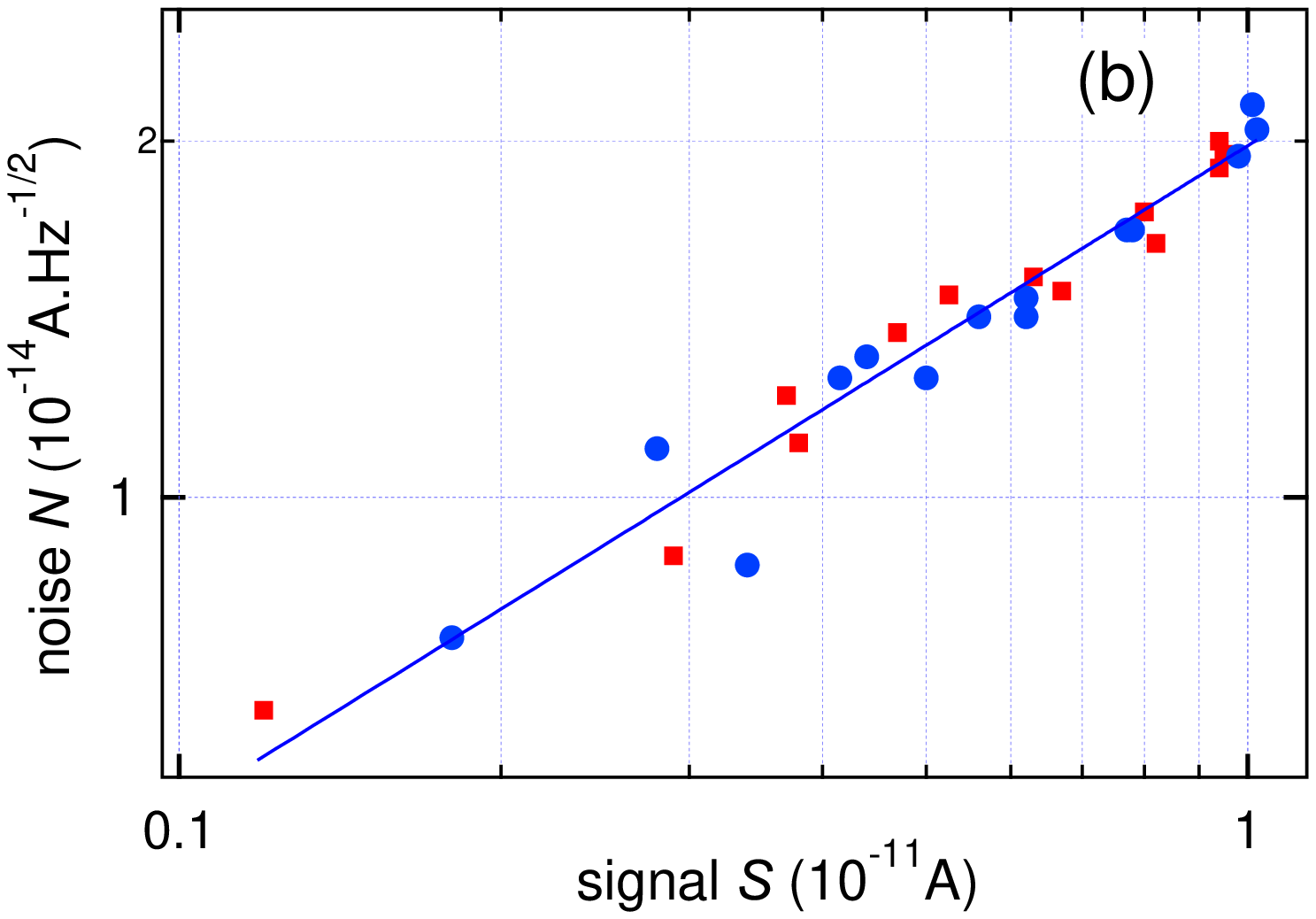}
\caption{Rms values of the noise $N$ of the fluorescence photodetector signal. (a) \textit{vs} the
Fourier frequency at intermediate atomic flux (upper trace) and without atomic flux (lower trace).
(b) \textit{vs} the contribution $S$ of the atomic fluorescence to the DC signal. The microwave
field is tuned to the centre of the central Ramsey fringe of the clock transition without phase
modulation. The atomic flux is varied by varying the power of the repumper beam (squares) or of the
cooling beams (bullets) in the 3D-optical molasses source. The straight line is a fit by a power
law with exponent 0.560~$\pm$~0.025.}\label{fig:snr}
\end{center}
\end{figure}

For the frequency stability measurements, the 10~MHz output of the VCXO locked to the fountain is
compared with the 10~MHz output of a maser contributing to the definition of TAI (maser 140-57-01)
using a frequency comparator (VCH-314). Fig.~\ref{fig:stab} displays a typical result obtained at
maximum flux. At 1~s, the phase noise of the maser dominates. For $\tau~>~200$~s, the Allan
deviation is consistent with 2$\times 10^{-13}\tau^{-1/2}$ and lies near the limit $\sim$ (1.5 to
1.8)$\times 10^{-13}\tau^{-1/2}$ associated with the atomic shot noise expected from the
signal-to-noise ratio discussed above. In addition, control measurements performed at half-maximum
flux
 were consistent with a degradation of the instability as $\phi^{-1/2}$, \textit{i.e.} with atomic
shot noise.

\begin{figure}[thb]
\begin{center}
\includegraphics[width=8cm]{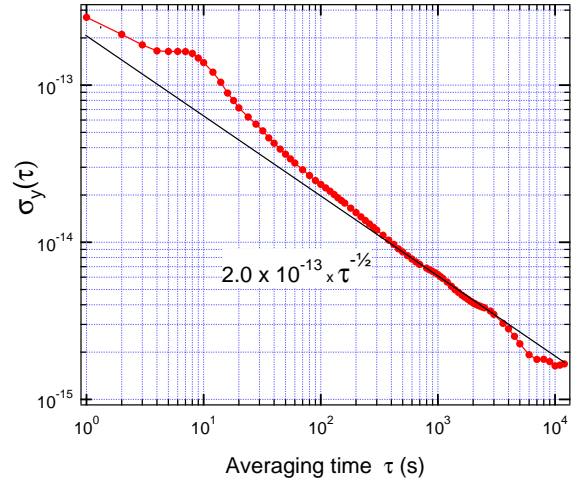}
\caption{Allan deviation $\sigma_{y}$ of the LO locked to FOCS-1 compared to the reference maser,
as a function of the averaging time $\tau$. The points represent the experimental data and the
continuous line indicates $2.0\times 10^{-13}~\tau^{-1/2}$.}\label{fig:stab}
\end{center}
\end{figure}

\section{The predicted intermodulation effect}\label{sec:dick}

\subsection{Allan deviation due to the intermodulation effect}
A theoretical analysis of a possible degradation of the frequency stability of a frequency standard
due to intermodulation effects has been presented in detail previously \cite{AJo1}. Here, only the
basic ingredients of the model and the main results will be recalled.

From a knowledge of the power spectral density (PSD) of the frequency fluctuations of the free
running oscillator $S_y^{LO}(f)$, the model aims at predicting the power spectral density
$S_y^{LLO}(f)$ of the oscillator locked to a Ramsey resonator using any modulation - demodulation
scheme. Though the treatment was developed specifically for the case of a continuous interrogation,
 it can also be applied to the pulsed case, or for that matter, to any-time dependence of the
atomic flux. From the PSD of the locked oscillator, the Allan deviation can then readily be
computed.

The spectrum of the LLO has two parts: the first corresponds to the error signal that controls the
LO frequency fluctuations; the second, not reduced by the loop gain, contains all \textit{even}
harmonics of the modulation frequency \fmod~and corresponds to a spurious signal generated by
down-conversion of the LO frequency fluctuations at harmonics of the modulation frequency. The
latter is the aliasing part of interest here and is given by Eq.(30) of Ref. \cite{AJo1}:

\begin{equation}
S_{y,Dc}^{LLO}(f) \simeq 2 \sum_{k=1}^{\infty} \frac{\vert c_{2k}\vert^{2}}{c_{0}^{2}}
\sinc^{2}(2k\pi f_{mod}T) S_{y}^{LO}(2k f_{mod})
\end{equation}
for Fourier frequencies $0 \leq f \leq f_{F}$, where the $c_{2k}$ are the Fourier coefficients of
the modulation-demodulation function, $T$ is an effective transit time between the two Ramsey
pulses (here $\simeq 0.49$~s) and $f_{F}$ is the bandwidth of the low-pass filter in the loop. A
remarquable property of this equation is that for the condition $T = \tmod/2$, \textit{i.e.} when
the modulation frequency is equal to the resonator linewidth $\Delta\nu_0$ ($\simeq$~1~Hz), all
terms in Eq.(1) are equal to zero: as a result the aliasing effect vanishes, whatever
modulation-demodulation scheme. The interpretation is a filtering effect of the LO frequency
fluctuations at the even harmonics of \fmod~by averaging over the transit time through the
resonator. In the particular case of square-wave phase modulation and square-wave demodulation the
$c_{2k}$ vanish for all values of $k$, which provides an even more robust cancellation effect.

The bandwidth of the frequency control loop $f_{F}$ is a fraction of a hertz. %, i.e. much smaller that f$_{cut}$.
Within this bandwidth, the frequency perturbations $S_{y,Dc}^{LLO}(f)$ due to the continuous
intermodulation effect can be considered as white noise (Eq. (1)). As a consequence, the Allan
deviation due to this effect is given by:

\begin{equation}
\sigma_{y,Dc}^2(\tau)= S_{y,Dc}^{LLO}(f=0)/2\tau
\end{equation}
provided the averaging time $\tau$ is longer than the time constant of the servo loop (typically
$\tau$ > 10~s).

For quantitative predictions, we must calculate the Fourier coefficients $c_{2k}$ involved in
Eq.(1)
%which depend on the modulation-demodulation scheme,
and use a model for $S_y^{LO}(f)$, the PSD of
the frequency fluctuations of the free LO. For $S_y^{LO}(f)$, the main sources of frequency
instability can be parametrised by the following expansion:
\begin{equation}
S_{y}(f)= \sum_{\alpha=-2}^{2} h_{\alpha} \times f^{\alpha}
\end{equation}
corresponding to random frequency ($h_{-2}$), flicker frequency ($h_{-1}$), white frequency
($h_{0}$), flicker phase ($h_{1}$) and white phase noise ($h_{2}$).

In Ref.~\cite{AJo2}, three usual types of modulation schemes were investigated: square-wave phase
modulation, sinusoidal-wave and square-wave frequency modulations, and, for each of them, either
first harmonic or wideband demodulation schemes were considered. The simulations of the
intermodulation effect were performed for a particular quartz oscillator exhibiting contributions
from $h_{-1}$ and $h_{1}$ types of noise, with a flicker floor Allan deviation of $3\times
10^{-13}$. The results show that for significant deviation (0.4 to 0.5~Hz) of \fmod~from the value
$\Delta\nu_0=1$~Hz (or its odd harmonics) the Allan deviation due to the intermodulation effect
reaches at most $10^{-13}$ for $\tau$ = 1~s, that is, a value below the present Allan deviation
limited by the atomic shot noise (see Sec. \ref{sec:perfos} where \fmod~is chosen equal to
$\Delta\nu_0=1$~Hz). Similar results were obtained with the three modulation schemes \footnote{We
note that differences between the different modulation schemes show up essentially in how the
effect vanishes in the vicinity of \fmod= $\Delta\nu_{0}$. Of course, for optimum performances of
the clock under normal operation these differences have to be considered and the best modulation
scheme adopted.}. Since our present oscillator is even better with a $\sim$~3 times lower flicker
phase noise, it is not well suited for demonstrating the intermodulation effect. It was thus
necessary to degrade deliberately the phase noise performance of the LO. We chose to inject white
phase noise ($h_{2}$-type) with an amplitude such that the intermodulation effect contribution is
$\sim$~2 times the atomic shot noise for an experimentally sensible deviation of \fmod~from 1~Hz
(or its odd harmonics).

\subsection{Prediction in the case of square-wave phase
modulation with white phase noise}

For square-wave phase modulation and wideband demodulation, the Fourier coefficients can be
calculated analytically, and from Eq.~(6.32) of Ref. \cite{AJo2} the $h_{2}$ contribution to the
intermodulation effect reads:
\begin{equation}
\sigma_{y,Dc}(\xi,\tau)= \frac{1}{2\pi^{2}\xi(1-\vert\xi-1\vert)}
\left\{\frac{h_{2}}{T^{2}}\,\xi^{2}\, g(\xi)\right\}^{1/2}\, \tau^{-1/2}
\end{equation}
where $ g(\xi)= \sum_{k=0}^{\infty} \sin^{4}(k\pi\xi)/k^{2}$ and $\xi$= \fmod/$\Delta \nu_{0}$. The
dependence of the Allan deviation on the frequency modulation \fmod~is shown in Fig. \ref{fig:Dick}
for $h_{2}=10^{-23}$ Hz$^{-3}$. Using the relation $S_\phi(f)$ = $S_{y}(f) \times (f_{0}/f)^{2}$
with $f_{0}$ $\simeq$ 9.2~GHz the clock frequency, this corresponds to a PSD of the phase noise
$S_\phi(f)$ of $10^{-3}$ rad$^{2}$~Hz$^{-1}$. One can see the notch effect of the continuous Ramsey
resonator at 1~Hz, the width $\Delta \nu_{0}$ of the resonance line, and the rapid degradation
either side of this value.

\begin{figure}[thb]
\begin{center}
\includegraphics[width=8.5cm]{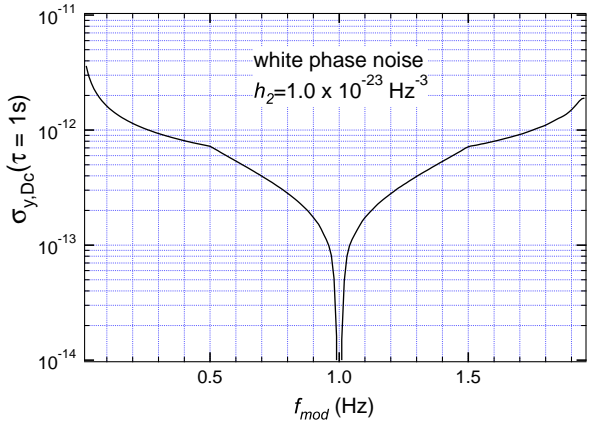}
\caption{Allan deviation due to the intermodulation effect using the model developed in \cite{AJo1}
for the case of square-wave phase modulation with white phase noise $h_{2}=10^{-23}$~Hz$^{-3}$, as
a function of the frequency modulation \fmod. Values calculated for an averaging time
$\tau$~=~1~s.}\label{fig:Dick}
\end{center}
\end{figure}

\section{Noise generation, measurements and results}\label{sec:measurements}

\begin{figure}[h]
%\begin{center}
\includegraphics[width=8.5cm]{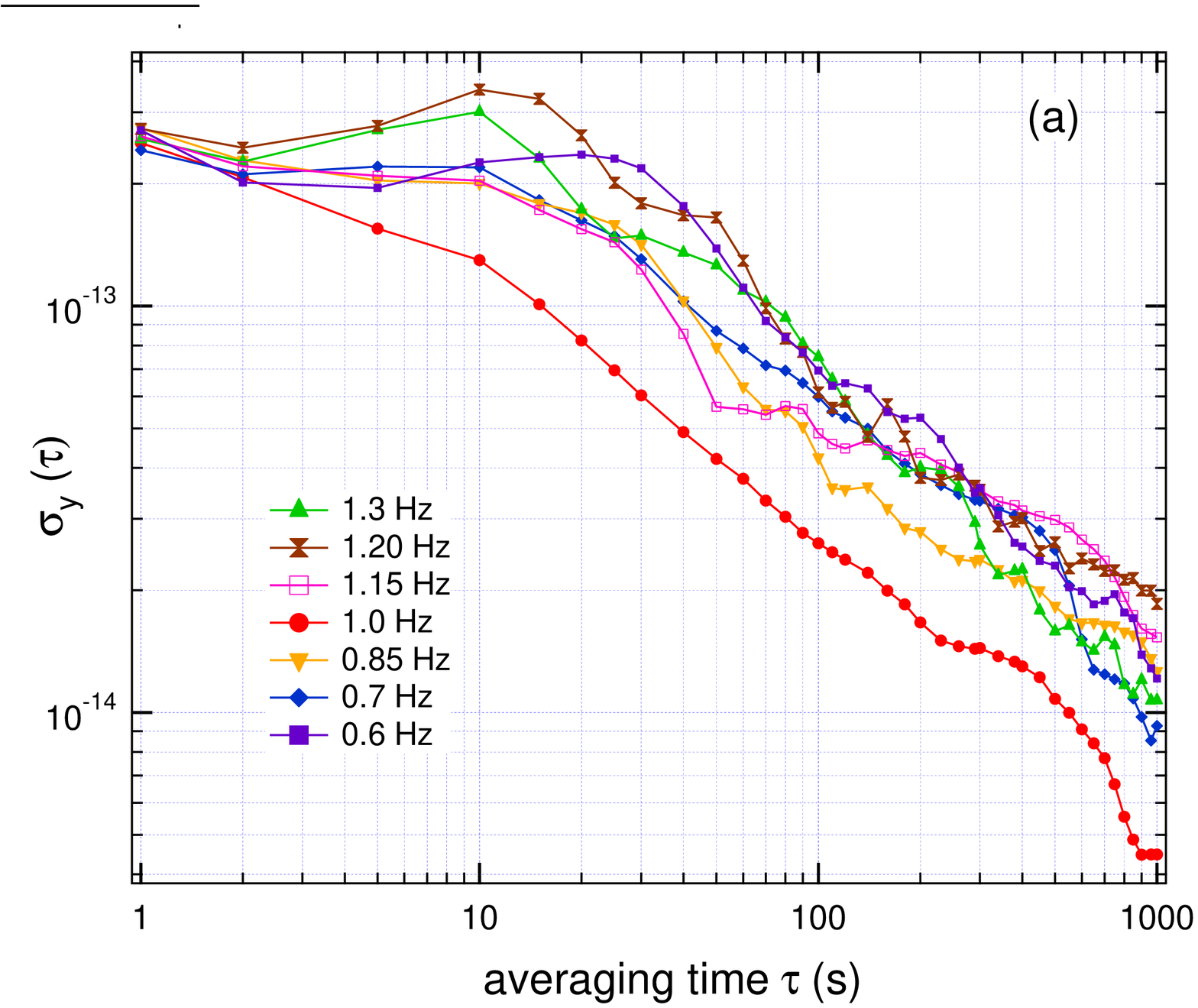}
\includegraphics[width=8.5cm]{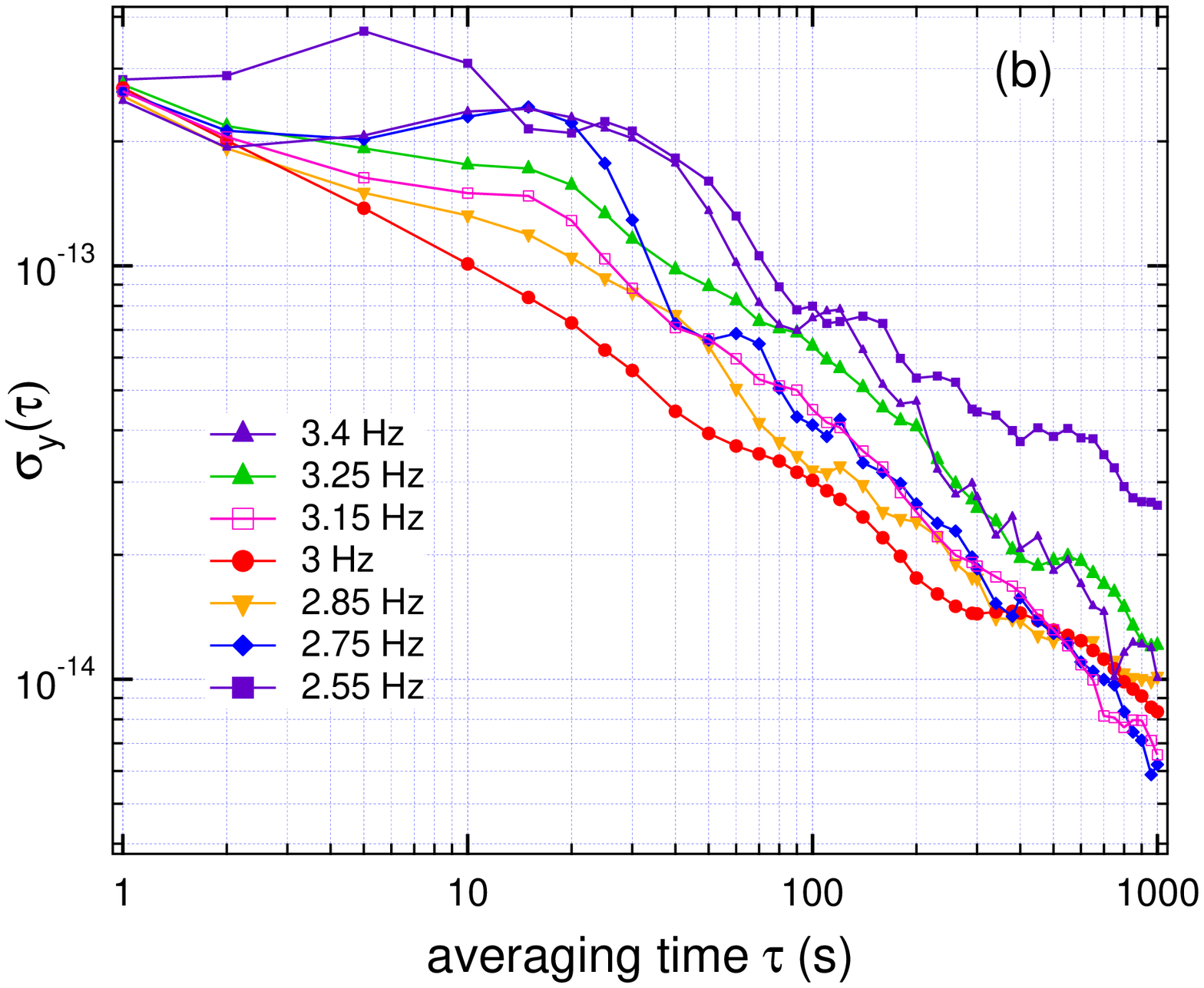}
\includegraphics[width=8.5cm]{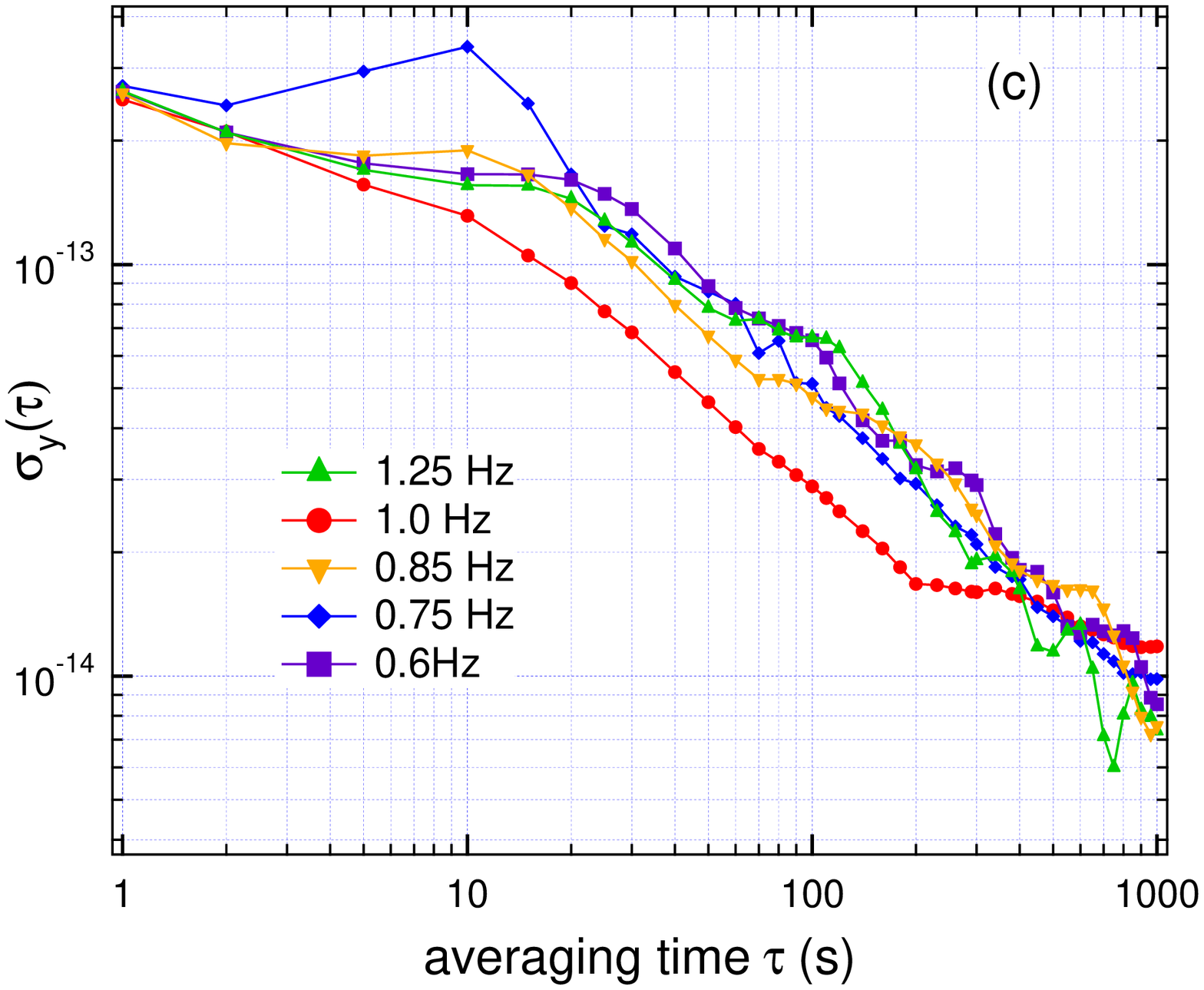}
\caption{Experimental Allan deviations of the LO locked to FOCS-1 with white phase noise injected
\textit{via} the synthesiser into the microwave field fed to the Ramsey resonator. (a) Square-wave
phase modulation with modulation frequency varied around 1~Hz, the (FWHM) linewidth of the
resonator. (b) Square-wave phase modulation with modulation frequency varied around 3~Hz. (c)
Square-wave frequency modulation with modulation frequency varied around 1~Hz.}\label{fig:Allan
Dev}
\end{figure}

To generate the phase noise in the microwave field, we add a noise voltage to the phase modulation
input of the 12.6~MHz generator (see Fig.
\ref{fig:loop}). % that drives the clock frequency synthesiser (see Sect. \ref{sec:loop}).
The noise voltage provided by a wideband white noise generator (DS340) is  low-pass filtered and
amplified beforehand. We adjust the cut-off frequency to select the first 20 harmonics of the
modulation frequency \fmod, since these give almost the whole contribution (>98\%) to the
intermodulation effect. The rms noise voltage is adjusted to provide a phase noise of order
$\sim$~30 mrad Hz$^{-1/2}$ ($h_{2}\approx 10^{-23}$~Hz$^{-3}$). Note the interest of injecting the
phase noise into the 12.6~MHz oscillation, rather than directly into the VCXO control voltage
input: the added frequency instability of the VCXO is then representative of the intermodulation
effect, and not of some parasitic effect due to a direct degradation of the VCXO.

Most of the frequency stability measurements were performed using square-wave phase modulation with
\fmod~varied around either 1~Hz or 3~Hz, \textit{i.e.} the first or third harmonic of the Ramsey
resonator linewidth. The Allan deviations, for each value of \fmod, obtained after a measurement
time of typically 2 hours are plotted in Fig. \ref{fig:Allan Dev}a and \ref {fig:Allan Dev}b,
respectively. For $\tau$ of a few hundred seconds, the VCXO is locked to the fountain and the Allan
deviations approach the $\tau^{-1/2}$ law (white frequency noise) expected from both atomic shot
noise and the intermodulation effect. For $\tau$~>~1000~s, the statistics are poorer and the values
less significant. We should note that, when the modulation frequency is changed, the slope of the
Ramsey discriminator also changes.
 For instance with \fmod~=~1.4~Hz, the slope is decreased by 30\% in square-wave phase modulation
mode \footnote{The phase of the demodulation has also to be readjusted, which is done by maximizing
the Ramsey slope when the microwave frequency is detuned by a half linewidth ($\pm$~0.5~Hz) from
the fringe centre.}. This means that even in the absence of the intermodulation effect, the
stability limited by the atomic shot noise can be degraded when \fmod~is varied. This effect has
been taken into account by our repeating the measurements without phase noise injected in the
synthesiser.
 The observed degradation due to the change of the discriminator slope agrees with that expected
 based on atomic shot noise and remains much smaller
 than the frequency instability observed when phase noise is injected (see below). Measurements were also
 performed using square-wave
 \textit{frequency} modulation. Simulations carried out in
 Ref. \cite{AJo2} indicate that the intermodulation effect should show up with the same order of magnitude.
 In this case the slope is smaller as is its dependence on \fmod. The experimental results are plotted in
 Fig. \ref{fig:Allan Dev}c.
In the three graphs of Fig. \ref{fig:Allan Dev}, the better
 stability obtained at \fmod~equal to 1 Hz (in (a) and (c)) or 3 Hz (in (b)) is quite conspicuous.

\begin{figure}[h]
\includegraphics[width=8.5cm]{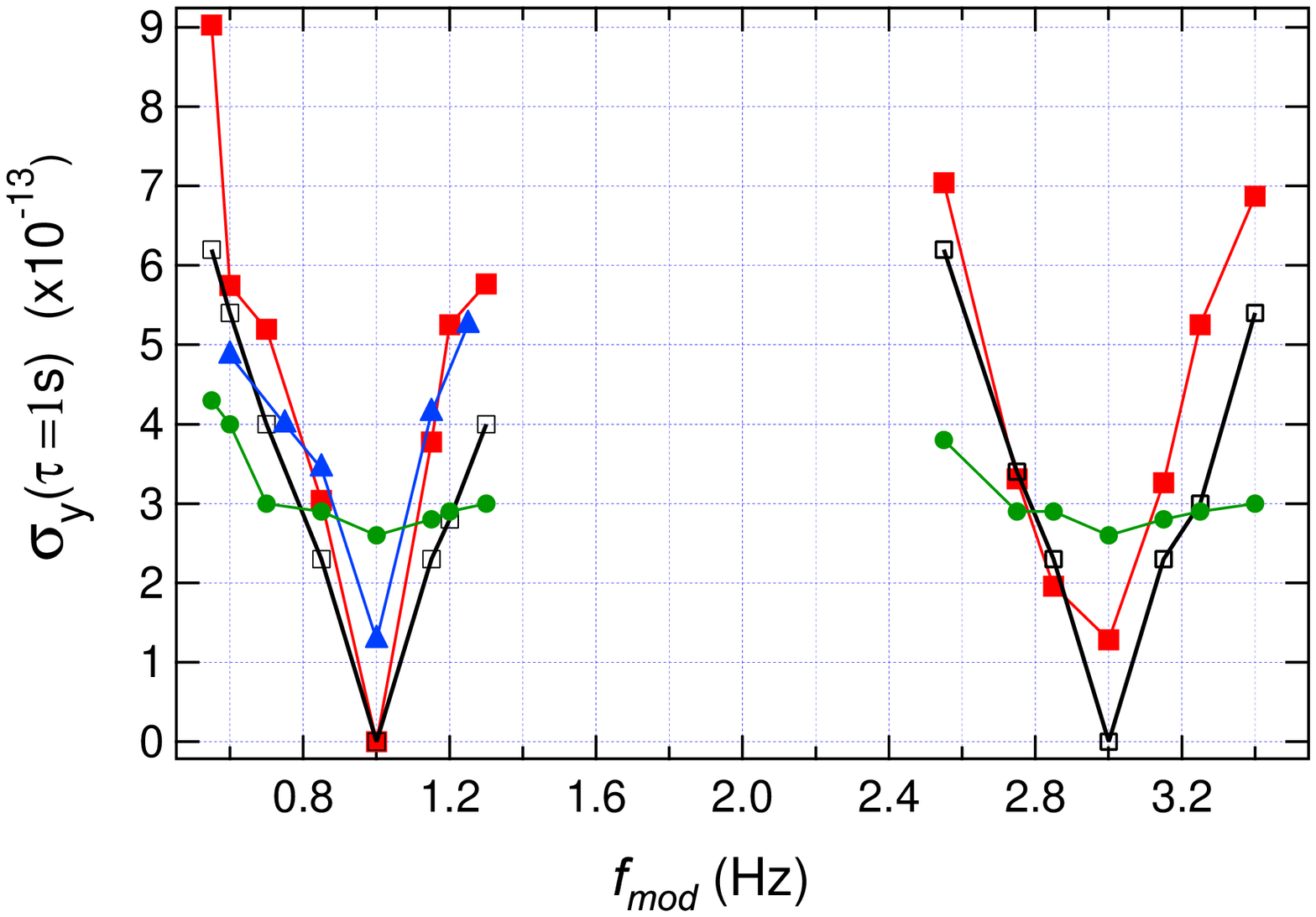}
\caption{Allan deviations \textit{vs} interrogation modulation frequency:  observed in square-wave
phase modulation without injected phase noise, used as references (bullets); observed with injected
phase noise after quadratic subtraction of the references (filled squares and triangles for
square-wave phase and square-wave frequency modulations, respectively); predicted intermodulation
effect for injected noise $h_{2}=10^{-23}$~Hz$^{-3}$ and square-wave modulation (open
squares).}\label{fig:results}
\end{figure}

With a view to extracting the intermodulation contribution to the measured instabilities and
comparing them with the prediction above, we subtracted quadratically the instabilities measured
without injected phase noise from the total instabilities
 measured in presence of phase noise. The values are taken for averaging times $\tau$ in the range
 100 to 400~s and
 scaled to 1~s using the $\tau^{-1/2}$ law. The results are plotted  as a function of \fmod ~in Fig. \ref{fig:results}. There
 is good agreement with the calculated intermodulation effect. The most important feature is the suppression
 of the effect when the modulation frequency equals an \emph{odd} harmonic of the Ramsey fringe linewidth.
Note that we have used the model of Ref. \cite{AJo1} which makes several simplifications, namely:
monokinetic atomic beam; infinitely short Rabi pulses, hence absence of transients; no phase shift
between modulation and demodulation waveforms. One might reasonably enquire whether these
idealizations could yield an underestimate of the effect. However,
 several points involved in the real experiment were considered in Ref. \cite{AJo2}: i) longitudinal velocity distribution of atoms
  (a few cm/s around the average launch velocity of 4~m/s, or longitudinal temperature of about 75~$\mu$K);
  ii) phase shift between modulation and demodulation; iii) introduction of dead times (blanking) of duration equal to the duration of the transients.
  From these simulations, the corresponding
  degradations of stability are expected to be very small compared with the atomic shot noise. Thus we believe
  that the observed
  slight excess of instability over the predicted value arises from some
  technical noise added in the generation of the phase noise \footnote{We note that for some values of \fmod,
  oscillations of the LO frequency with a long period (in the range 100 to 2000~s) appeared. However, they
  were not reproduced with a second noise generator (DS345) whose algorithm to generate random noise is different. This
highlights just how hard it is to generate perfectly white noise without adding side effects in the
loop.}. We recall that the latter corresponds to an
  increase by 2 to 3 orders of magnitude with respect to the undegraded phase noise,
  and to an rms phase noise of 2\% of the $\pi/2$ p.p. phase modulation.
  Concerning the possible effect of transients, we note that these could show up when \fmod~is varied around 3~Hz instead of 1~Hz.
  In fact, no significant degradation is observed (see Fig. \ref{fig:results}).

\section{Conclusion}
In this paper, we show that a continuous cold-atom fountain clock that uses a commercial quartz
crystal oscillator has a frequency stability  limited mainly by atomic shot noise. By deliberately
degrading the noise spectrum of the microwave field fed to the Ramsey cavity, we have verified the
model specifically developed to predict the intermodulation effect in a continuous interrogation
scheme, in particular the cancellation of this effect when the modulation frequency involved in the
microwave frequency locking loop equals an odd harmonic of the resonator linewidth (FWHM
$\simeq$~1~Hz). The magnitude of the effect when the modulation frequency is changed from these
values is observed at the level predicted by the model. This experimental verification is important
because several points involved in the actual experiment, \emph{e.g.} transients effects, are
difficult to include in the model. To observe the intermodulation effect at a level comparable with
that of the present atomic shot noise, we were led to degrade the noise of our quartz oscillator by
2 to 3 orders of magnitude. This work thus demonstrates the advantage of the continuous fountain
approach in reducing intermodulation-related local oscillator noise and opens the possibility of
improving significantly the short-term stability of a continuous fountain by increasing the atomic
flux.

\section{Acknowledgments}
We wish to thank M.D. Plimmer for critical reading of the manuscript. METAS acknowledges the Centre
National de la Recherche Scientifique (CNRS) for having made possible the stay of J.G. at its
institute.

\end{document}